\documentclass[12pt]{article}
\usepackage{ulem}
\usepackage{amssymb,amsmath,comment,mathtools}
\usepackage[dvipdfmx]{graphicx}
\usepackage{subfigmat}
\usepackage{braket}
\usepackage{hyperref} 	  		
\usepackage{epsfig}
\usepackage{here}
\usepackage{color}
\usepackage{bm}

\newcommand{\ba}{\begin{align}}
\newcommand{\ea}{\end{align}}

\def\nn{\nonumber}

\def\bea{\begin{eqnarray}}
\def\eea{\end{eqnarray}}
 \makeatletter
\def\alt{\mathrel{\mathpalette\gl@align<}}
\def\agt{\mathrel{\mathpalette\gl@align>}}
\def\gl@align#1#2{\lower.6ex\vbox{\baselineskip\z@skip\lineskip\z@
\ialign{$\m@th#1\hfil##\hfil$\crcr#2\crcr\sim\crcr}}} \makeatother

\usepackage{feynmp} 
\usepackage{tikz-feynman}
\tikzfeynmanset{compat=1.1.0}

\usepackage{tikz-feynhand}

\begin{document}
\begin{flushleft}
\end{flushleft}

\vspace*{1.0cm}
\begin{center}
\baselineskip 20pt 
{\Large\bf 
Non-thermal baryogenesis \\
 from MSSM flat direction}
\vspace{1cm}

{\large 
Naoyuki Haba${}^{b,c}$, Yasuhiro Shimizu${}^{b,c}$, Yoshihiro Tanabe${}^{a,b}$, \\
and Toshifumi Yamada${}^{d}$
} \vspace{.5cm}

{\baselineskip 20pt \it
${}^{a}$Institute of Science and Engineering, Shimane University, Matsue 690-8504, Japan\\
${}^{b}$Department of Physics, Osaka Metropolitan University, Osaka 558-8585, Japan \\
${}^{c}$Nambu Yoichiro Institute of Theoretical and Experimental Physics (NITEP),
Osaka Metropolitan University, Osaka 558-8585, Japan\\
${}^{d}$Institute for Mathematical Informatics, Meiji Gakuin University, Yokohama 244-8539, Japan
}

\vspace{1.5cm} {\bf Abstract} \end{center}
\noindent

We study an inflection point inflation scenario where a flat direction of the minimal supersymmetric standard model (MSSM) is identified with the inflaton.
We focus on the case where the flat direction (inflaton) has non-zero baryon number, and consider a non-thermal baryogenesis scenario
 where the decay of the inflaton at the reheating directly generates baryon asymmetry of the Universe.
Specifically, we consider a $udd$ flat direction that is lifted by a superpotential operator of dimension 6,
 and show that inflection point inflation with the $udd$ flat direction 
 can be compatible with cosmological observations and can account for the baryon asymmetry of the Universe.

\section{Introduction}

The nature of the inflaton that drove the primordial inflation~\cite{Guth:1980zm} and the origin of the matter-antimatter asymmetry of the Universe~\cite{FD2},
 are two major mysteries of particle physics and cosmology.

A lot of inflation scenarios and inflaton candidates have been proposed. One of the attractive scenarios is so-called inflection point inflation~\cite{Allahverdi:2006iq}-\cite{Choudhury}, since it easily satisfies the constraints from cosmological observations.
Also, supersymmetric (SUSY) extension of the standard model is a viable candidate for physics above TeV scale, because SUSY can stabilize the large hierarchy between the Planck scale and the electroweak scale.
The minimal SUSY standard model (MSSM) exhibits an interesting property that certain combinations of scalar fields have vanishing triple and quartic couplings~\cite{FD1}.
Such combinations are called ``flat directions".
The flat directions are a natural candidate for the inflaton of inflection point inflation~\cite{Allahverdi:2006iq}-\cite{Allahverdi:2006we}.

In this paper, we study the scenario where a flat direction in MSSM is identified with the inflaton that induces quasi-inflection point inflation. 
In this model, the first derivative of the inflaton potential is small but non-zero value, which allows the scalar spectral index $n_s$ to be within the observed range unlike in a bona fide inflection point inflation model ~\cite{Allahverdi:2006iq,qIPI}.
We concentrate on the case where the inflaton, which is a MSSM flat direction, has non-zero baryon number, and investigate non-thermal baryogenesis from its decay at the reheating.

Specifically, we consider the case where a $udd$ flat direction is lifted by a superpotential operator of dimension 6, which can be compatible with the Planck / BICEP data and multi-TeV squark masses~\cite{qIPI}. We present a benchmark parameter set with sparticle masses in the range of 2000GeV to 4000GeV that successfully explains the observed scalar power spectrum amplitude $P_\zeta (k_*)$ and the scalar spectral index $n_s$, satisfies the bound on tensor-to-scalar ratio $r$, and moreover explains the observed baryon number density through non-thermal baryogenesis.

\vspace{1em}
This paper is organized as follows:
In Section 2, we describe the model where MSSM is extended by the dimension 6 operator that lifts the $udd$ flat direction, and derive the potential for this direction that can realize inflection point inflation.
In Section 3, we construct a quasi-inflection point inflation model by introducing a slight deviation $\alpha$ to the mass relation equation for a bona fide inflection point inflation that leads to a non-zero first derivative of the potential at the inflection point. We then derive the expressions for the slow-roll parameters $\eta$, $\epsilon$, the scalar power spectrum amplitude $P_\zeta (k_*)$, and the Hubble rate during and at the end of inflation.
In Section 4, the reheating of the Universe through the decay of the radial component of the $udd$ direction, which is identified with the inflaton, is investigated.
In Section 5, we show that the decay mode of the flat direction decaying into a quark/antiquark and a Higgsino generates baryon number asymmetry,
 and explicitly calculate the CP asymmetry parameter and the baryon number yield.
In Section 6, a benchmark parameter set is shown which meets the constraints on cosmological observables and accounts for the baryon number of the Universe. 
Section 7 concludes the paper.

\section{Model}

The superpotential of the model is given by
\begin{align}
W = W_{\rm MSSM} + \frac{\lambda}{2 M_p^{3}}(U^c_iD^c_jD^c_k)^2,
\label{nr}
\end{align}
 where $W_{\rm MSSM}$ denotes the superpotential of MSSM, and the second term is a higher-dimensional term.
Here $U^c,D^c$ respectively denote the isospin-singlet up-type and down-type quark superfields, $i,j,k$ are flavor indices,
 $\lambda$ is a coupling constant that is taken to be real positive, $M_p$ is the reduced Planck mass (2.44 $\times 10^{18}$ GeV),
 and the color indices are summed in the bracket $(...)$.
Additionally, the squarks possess soft SUSY breaking masses and $A$-term proportional to the higher-dimensional term as
\begin{align}
V_{\rm soft} \ \supset \ m_{\tilde{u}_i}^2 \tilde{u}_i^{c \dagger} \tilde{u}_i^c + m_{\tilde{d}_j}^2 \tilde{d}_j^{c \dagger} \tilde{d}_j^c + m_{\tilde{d}_k}^2 \tilde{d}_k^{c \dagger} \tilde{d}_k^c - {\cal A}\frac{\lambda}{2 M_p^{3}} (\tilde{u}_i^c\tilde{d}_j^c\tilde{d}_k^c)^2 + {\rm h.c.},
\end{align}
 where $\tilde{u}^c,\tilde{d}^c$ respectively denote the scalar components of $U^c,D^c$.

We consider a $udd$ flat direction in MSSM, given by
\begin{equation}
(\tilde{u}_i^c)^\alpha=\frac{1}{\sqrt{3}}\Phi,\;
(\tilde{d}_j^c)^\beta=\frac{1}{\sqrt{3}}\Phi,\;
(\tilde{d}_k^c)^\gamma=\frac{1}{\sqrt{3}}\Phi,
\end{equation}
 where $\alpha, \beta, \gamma$ are color indices, and $\Phi$ is a complex scalar field that parametrizes the flat direction.
The flatness constraints require $\alpha \neq \beta \neq \gamma \neq \alpha$ and $j \neq k$.
The potential for the flat direction reads
\begin{equation}
    V(\Phi)=m^2_\Phi |\Phi|^2 - \mathcal{A}\dfrac{\lambda}{54 M^{3}_p}\Phi^6 - {\rm h.c.} + \frac{\lambda^2}{81M^{6}_p}|\Phi|^{10},
\end{equation}
 where $m^2_\Phi=(m^2_{\tilde{u}_i} + m^2_{\tilde{d}_j} + m^2_{\tilde{d}_k})/3$.
We rewrite the flat direction as $\Phi=\phi e^{i\,\theta}/\sqrt{2}$, where $\phi$ denotes the radial component that is real.
Then the potential is recast into
\begin{equation}
    V(\Phi)=V(\phi,\theta)=\dfrac{m^2_\Phi}{2} \phi^2 - |\mathcal{A}|\dfrac{\lambda}{216 M^3_p} \phi^6 \cos(6\theta+\theta_A) + \dfrac{\lambda^2}{2592M^{6}_p}\phi^{10}, 
\end{equation}
 where $\theta_A$ is the phase of ${\cal A}$.
This potential is minimized for $\theta$ satisfying $\cos(6\theta+\theta_A)=1$. When $\theta$ is stabilized at such a value, the potential for $\phi$ becomes
\begin{equation}
    V(\phi)=\dfrac{m^2_\Phi}{2} \phi^2 - |\mathcal{A}|\dfrac{\lambda}{216 M^3_p} \phi^6 + \dfrac{\lambda^2}{2592M^{6}_p}\phi^{10}.
\label{potential}
\end{equation}

We assume that the SUSY particle masses have the following hierarchy:
\begin{align}
{\rm (Higgsino \ and \ gaugino \ masses)} < {\rm (isospin \ singlet \ squark \ masses)} < {\rm (isospin \ doublet \ squark \ masses)}
\label{massspectrum}
\end{align}
As a result, the squarks that constitute the flat direction decay into a gaugino$+$a quark/antiquark and a Higgsino$+$a quark/antiquark.
The latter decay mode realizes non-thermal baryogenesis.
The parameters relevant to non-thermal baryogenesis are the quark Yukawa couplings, $\mu$-term and quark A-terms, defined as follows:
\begin{align}
W_{\rm MSSM} \ &\supset \ y^u_{ji} \ Q_j H_u U^c_i + y^d_{ji} \ Q_j H_d D^c_i + \mu \ H_u H_d,
\\
V_{\rm soft} \ &\supset \ A^u_{ji} \ \tilde{q}_j H_u \tilde{u}^c_i + A^d_{ji} \ \tilde{q}_j H_d \tilde{d}^c_i,
\end{align}
 where $Q$ denotes the isospin-doublet quark superfields, $\tilde{q}$ their scalar components, and
$H_u,H_d$ represent both the Higgs superfields and their scalar components.
\\

\section{Inflection point inflation}

In the rest of the paper, we focus on the case with $9|{\cal A}|/(10\lambda M_p) <1$.
\\

To realize the inflection point inflation, the inflaton potential Eq.~(\ref{potential}) should have a quasi-inflection point.
The condition for the existence of a quasi-inflection point is
\begin{align}
m_\Phi^2 = \frac{|{\cal A}|^2}{20}(1+\alpha),
\label{condition}
\end{align}
 where $|\alpha | \ll 1$.
We define the quasi-inflection point, $\phi=\phi_0$, as the point satisfying $V''(\phi_0)=0$ and $V'(\phi_0) \propto \alpha$.
Also, without loss of generality, we consider only the positive quasi-inflection point $\phi_0 > 0$.
Given Eq.~(\ref{condition}), it is obtained as
\begin{align}
\phi_0 = \sqrt{2}\left(\frac{9|{\cal A}|M_p^3}{10\lambda}\right)^{\frac{1}{4}}\left(1-\frac{\alpha}{32}\right) + O(\alpha^2).
\end{align}
The potential and its first and third derivatives at $\phi=\phi_0$ read
\begin{align}
V(\phi_0) = \frac{2}{75}  |{\cal A}|^2  \left(\frac{9|{\cal A}|M_p^3}{10\lambda}\right)^{\frac{1}{2}} + O(\alpha),
\label{v0}\\
V'(\phi_0) = \alpha\frac{1}{10\sqrt{2}}  |{\cal A}|^2  \left(\frac{9|{\cal A}|M_p^3}{10\lambda}\right)^{\frac{1}{4}} + O(\alpha^2),
\\
V'''(\phi_0) = \frac{8}{5\sqrt{2}}  |{\cal A}|^2  \left(\frac{9|{\cal A}|M_p^3}{10\lambda}\right)^{-\frac{1}{4}} + O(\alpha).
\label{v03}
\end{align}
We restrict ourselves to the case with $\alpha>0$ so that $V'(\phi_0)>0$ holds.


In the inflection point inflation, $\phi$ slow-rolls from the vicinity of the quasi-inflection point,
 to $\phi=\phi_{\rm end}$ at which the slow-roll condition is violated and inflation ends.
The number of e-folds as a function of $\phi$ is calculated as
\begin{align}
N(\phi) &= \frac{1}{M_p^2}\int_\phi^{\phi_{\rm end}} {\rm d}\phi \frac{V(\phi)}{-V'(\phi)}
\simeq \frac{1}{M_p^2}\int_\phi^{\phi_{\rm end}} {\rm d}\phi \frac{V(\phi_0)}{-V'(\phi_0)-\frac{1}{2}(\phi-\phi_0)^2V'''(\phi_0)}
\nn\\
&=N_0 \left[ \arctan\left(\frac{N_0 \ \eta(\phi)}{2}\right) - \arctan\left(\frac{N_0 \ \eta(\phi_{\rm end})}{2}\right) \right],
\label{nefolds}
\end{align}
 where $N_0$ is a quantity defined and calculated as
\begin{align}
N_0 \equiv \frac{1}{M_p^2}\sqrt{\frac{2V(\phi_0)^2}{V'(\phi_0)V'''(\phi_0)}}
= \frac{2}{15}\alpha^{-\frac{1}{2}}\left(\frac{9|{\cal A}|}{10\lambda M_p}\right)^{\frac{1}{2}}.
\end{align}
Since the term inside $[...]$ in Eq.~(\ref{nefolds}) is smaller than $\pi$, we need $N_0 \gg1$ to have a sufficient number of e-folds.
Hence, $\alpha$ should be fine-tuned as
\begin{align}
\alpha \ll \frac{9|{\cal A}|}{10\lambda M_p}.
\end{align}

The slow-roll parameters as functions of $\phi$ are calculated as
\begin{align}
\eta(\phi) &= M_p^2\frac{V''(\phi)}{V(\phi)} \simeq M_p^2\frac{V'''(\phi_0)}{V(\phi_0)}(\phi-\phi_0) 
= 30\sqrt{2}\left(\frac{10\lambda}{9|{\cal A}|M_p^{\frac{1}{3}}}\right)^{\frac{3}{4}}(\phi-\phi_0),
\label{eta}\\
\varepsilon(\phi) &= \frac{M_p^2}{2}\left(\frac{V'(\phi)}{V(\phi)}\right)^2 
\simeq \frac{M_p^2}{2}\left(\frac{V'(\phi_0)+\frac{1}{2}(\phi-\phi_0)^2V'''(\phi_0)}{V(\phi_0)}\right)^2
\nn\\
&= \frac{1}{14400}\left(\frac{9|{\cal A}|}{10\lambda M_p}\right)^{\frac{3}{2}}\left(\frac{4}{N_0^2} + \eta(\phi)^2\right)^2.
\label{epsilon}
\end{align}
Using the above results, the scalar power spectrum amplitude, the scalar spectral index and the tensor-to-scalar ratio at the pivot scale $k_*$ are computed as
\begin{align}
P_\zeta(k_*) &= \frac{V(\phi_*)}{24\pi^2 M_p^4 \varepsilon(\phi_*)} \simeq \frac{V(\phi_0)}{24\pi^2 M_p^4 \varepsilon(\phi_*)}
= \frac{16}{\pi^2}\frac{10\lambda |{\cal A}|}{9M_p}\left(\frac{4}{N_0^2} + \eta(\phi_*)^2\right)^{-2},
\end{align}
 and $n_s = 1 - 6\varepsilon(\phi_*) + 2\eta(\phi_*)$ and $r= 16 \varepsilon(\phi_*)$,
 where $\phi_*$ is the inflaton VEV when the pivot scale exited the horizon.

Let us determine $\phi_{\rm end}$.
The slow-roll condition is violated when $|\eta(\phi)|=1$ or $\varepsilon(\phi)=1$ holds.
Eqs.~(\ref{eta}),(\ref{epsilon}) give that
 $|\eta(\phi)|=1$ holds before $\varepsilon(\phi)=1$ holds because $9|{\cal A}|/(10\lambda M_p) <1$ and $N_0\gg1$.
Hence $\phi_{\rm end}$ is determined by the relation $\eta(\phi_{\rm end})=-1$, which yields
\begin{align}
\phi_{\rm end} - \phi_0 = -\frac{1}{30\sqrt{2}}\left(\frac{9|{\cal A}| M_p^{\frac{1}{3}}}{10\lambda}\right)^{\frac{3}{4}}.
\end{align}
Note that $|\phi_{\rm end} - \phi_0|$ is much smaller than $\phi_0$ because there holds
\begin{align}
\frac{|\phi_{\rm end} - \phi_0|}{\phi_0} \simeq \frac{1}{60}\left(\frac{9|{\cal A}|}{10\lambda M_p}\right)^{\frac{1}{2}} \ll 1.
\end{align}
As a result, the Hubble rate at the end of inflation, $H(\phi_{\rm end})$, is approximated by that during inflation, $H_{\rm inf}$,
 and is calculated as
\begin{align}
H(\phi_{\rm end}) \simeq H_{\rm inf} \simeq \sqrt{\frac{V(\phi_0)}{3M_p^2}}
= \frac{\sqrt{2}}{15}|{\cal A}| \left(\frac{9|{\cal A}|}{10\lambda M_p}\right)^{\frac{1}{4}}.
\label{hubble}
\end{align}
\\

\section{Reheating}
\label{reheating}

The decay of the radial component of the flat direction $\phi$ reheats the Universe.
To gain insight into the process of reheating,
 we compare the Hubble rate at the end of inflation $H(\phi_{\rm end})$ with the total width of $\phi$.

For simplicity, hereafter we focus on the case where the squarks comprising the flat direction are the superpartners of light flavor quarks.
Since their Yukawa couplings are small, and given the SUSY particle mass spectrum  hierarchy in Eq.~(\ref{massspectrum}),
 the main decay channel of $\phi$ is the decay into a gluino and a quark/antiquark.
Thus, the total width of $\phi$, denoted by $\Gamma_\phi$, is estimated as
\begin{align}
\Gamma_\phi &\sim \frac{1}{6}\left\{
\Gamma(\tilde{u}^c_i \to u_i^\dagger \tilde{g})
+\Gamma(\tilde{d}^c_j \to d_j^\dagger \tilde{g})
+\Gamma(\tilde{d}^c_k \to d_k^\dagger \tilde{g}) \right.
\nn\\
&\left.\ \ \ \ \ \  + \Gamma(\tilde{u}^{c\dagger}_i \to u_i \tilde{g})
+\Gamma(\tilde{d}^{c\dagger}_j \to d_j \tilde{g})
+\Gamma(\tilde{d}^{c\dagger}_k \to d_k \tilde{g})\right\}
\nn\\
&= \frac{1}{48\pi}
\left\{ m_{\tilde{u}_i}\left(1-\frac{M_{\tilde{g}}^2}{m_{\tilde{u}_i}^2}\right)^2
+ m_{\tilde{d}_j}\left(1-\frac{M_{\tilde{g}}^2}{m_{\tilde{d}_j}^2}\right)^2
+ m_{\tilde{d}_k}\left(1-\frac{M_{\tilde{g}}^2}{m_{\tilde{d}_k}^2}\right)^2
 \right\}\frac{8g_s^2}{3},
 \label{totalwidth}
\end{align}
 where $M_{\tilde{g}}$ denotes the gluino mass and $g_s$ the QCD gauge coupling. Here the quark masses are neglected.

If there is no hierarchy among $m_{\tilde{u}_i},m_{\tilde{d}_j},m_{\tilde{d}_k}$, the total width is approximated by
\begin{align}
\Gamma_\phi \sim \frac{m_\Phi}{16\pi}\left(1-\frac{M_{\tilde{g}}^2}{m_\Phi^2}\right)^2\frac{8g_s^2}{3},
\label{gamma}
\end{align}
 where $m_\Phi$ is the mass of the flat direction.
On the other hand, $H(\phi_{\rm end})$ in Eq.~(\ref{hubble}) can be rewritten with $m_\Phi$ through Eq.~(\ref{condition}) as
\begin{align}
H(\phi_{\rm end}) \simeq \frac{2\sqrt{10}}{15} m_\Phi \left(\frac{9|{\cal A}|}{10\lambda M_p}\right)^{\frac{1}{4}}.
\label{hubble2}
\end{align}

If $9|{\cal A}|/(10\lambda M_p)$ is sufficiently small such that 
\begin{align}
\frac{1}{16\pi}\left(1-\frac{M_{\tilde{g}}^2}{m_\Phi^2}\right)^2\frac{8g_s^2}{3}
 \gg \frac{2\sqrt{10}}{15} \left(\frac{9|{\cal A}|}{10\lambda M_p}\right)^{\frac{1}{4}},
\label{condition2}
\end{align}
 then the total width of $\phi$ far exceeds the Hubble rate.
It follows that cosmic expansion during the process of reheating is negligible
 and reheating temperature $T_R$ satisfies, by energy conservation,
\begin{align}
\frac{\pi^2}{30}g_{\rm eff} T_R^4 &\simeq 3M_p^2 H(\phi_{\rm end})^2,
\label{reheating}
\end{align}
where $g_{\rm eff}$ is the effective relativistic degree of freedom at temperature $T=T_R$.
Also, the scale factor at the reheating is approximated by that at the end of inflation.
\\

\section{Non-thermal baryogenesis}
\label{nonthermal}

$\phi$ also decays into a Higgsino$+$a quark/antiquark,
 but the branching ratio is subdominant.
However, this decay mode gives rise to the baryon number asymmetry of the Universe through non-thermal baryogenesis, as we discuss below.

There is a small difference in the partial widths of $\phi$ decaying into a Higgsino and a quark, and a Higgsino and an antiquark.
This difference comes from the combination of CP phase and strong phase~\cite{SP}.
Here, the CP phase is provided by those of squark A-terms and $\mu$-term,
 while the strong phase is provided by the 1-loop diagrams involving squark-Higgs boson-quark loop and quark-Higgsino-squark loop depicted in Fig.1.

\begin{figure}
 \centering
    \includegraphics[width=0.85\linewidth]{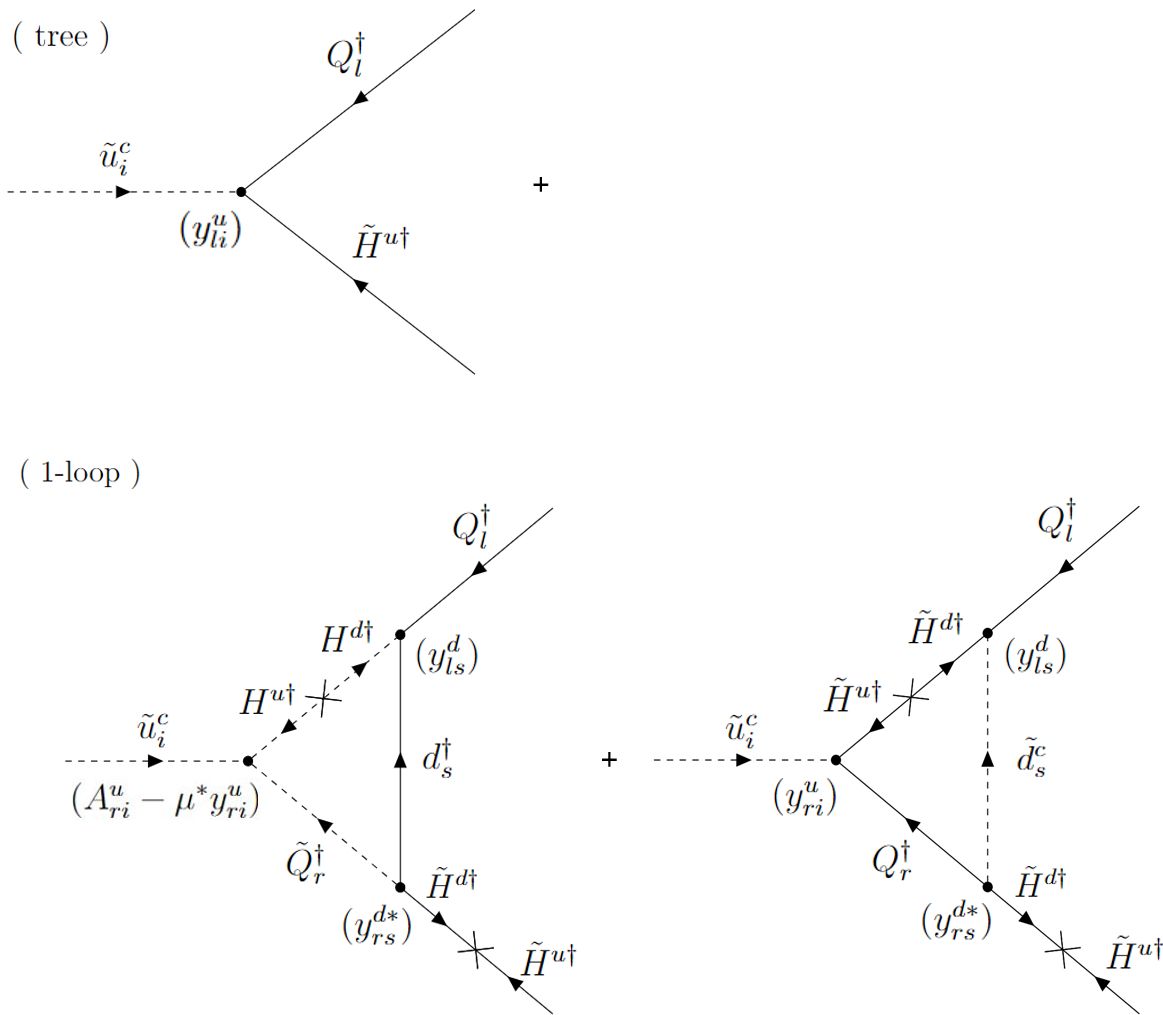}
    \caption{The Feynman diagram for the decay case : $\tilde{u}^c_i$ $\rightarrow$ $Q^{u\dag}_l$ + $\tilde{H}^{0\dag}$ }
\end{figure}

Let us denote by $Q_l^u,Q_l^d$ the up-type and down-type isospin-doublet quarks with flavor $l$,
 and denote by $\tilde{H}^0,\tilde{H}^\pm$ the neutral and charged components of Higgsinos.
The difference between the $\tilde{u}^c_i \to Q_l^{u\dagger} + \tilde{H}^{0\dagger}$ and $\tilde{u}_i^{c \dagger} \to Q_l^u + \tilde{H}^0$ partial widths
 is calculated by the following formula~\cite{SP1,SP4} (we have analogous formulas for the decays involving a chargino and the decays of $\tilde{d}^c_i$):
\begin{align}
&\Gamma(\tilde{u}^c_i \to Q_l^{u\dagger} + \tilde{H}^{0\dagger}) - \Gamma(\tilde{u}_i^{c \dagger} \to Q_l^u + \tilde{H}^0)
\nn\\
&=\frac{1}{16\pi m_{\tilde{u}^i}}\left(1-\frac{|\mu|^2}{m_{\tilde{u}^i}^2}\right)|\mu|
\left[\sum_{r=1}^3
2i\left\{ y^{u *}_{li} \ (y^d y^{d\dagger})_{lr} (A^u_{ri}-\mu^* y^u_{ri}) - y^{u}_{li} \ (y^d y^{d\dagger})_{lr}^* (A^{u*}_{ri}-\mu\, y^{u*}_{ri}) \right\} \right.
\nn\\
&\times\left\{ \int\frac{{\rm d}^4\ell}{(2\pi)^4}(2q\cdot \ell-2q^2) (-i\pi)^2\delta\left(\ell^2 - m_H^2\right)\delta\left((\ell-q)^2\right)P\frac{1}{(\ell-p)^2-m_{\tilde{Q}_r}^2} \right.
\label{integral1}\\
&\ \ \ \ \ \ \ \ + \int\frac{{\rm d}^4\ell}{(2\pi)^4}(2q\cdot \ell-2q^2) (-i\pi)^2 P\frac{1}{\ell^2 - m_H^2}\delta\left((\ell-q)^2\right)\delta\left((\ell-p)^2-m_{\tilde{Q}_r}^2\right)
\label{integral2}\\
&\left. \ \ \ \ \ \ \ \ + \int\frac{{\rm d}^4\ell}{(2\pi)^4}(2q\cdot \ell-2q^2) (-i\pi)^2\delta\left(\ell^2 - m_H^2\right)P\frac{1}{(\ell-q)^2}\delta\left((\ell-p)^2-m_{\tilde{Q}_r}^2\right)
\right\}
\label{integral3}\\
&+\sum_{s=1}^3
2i\left\{ y^{u *}_{li} \ y^d_{ls} \ (y^{d\dagger}y^u)_{si} \  \mu -  y^{u}_{li} \ y^{d*}_{ls} \ (y^{d\dagger}y^u)_{si}^* \ \mu^* \right\}
\nn\\
&\times\left\{ \int\frac{{\rm d}^4\ell}{(2\pi)^4}(-2q\cdot \ell + 2q\cdot p) (-i\pi)^2\delta\left(\ell^2 - |\mu|^2\right)\delta\left((\ell-q)^2-m_{\tilde{d}_s}^2\right)P\frac{1}{(\ell-p)^2} \right.
\label{integral4}\\
&\ \ \ \ \ \ \ \ + \int\frac{{\rm d}^4\ell}{(2\pi)^4}(-2q\cdot \ell + 2q\cdot p) (-i\pi)^2 P\frac{1}{\ell^2 - |\mu|^2}\delta\left((\ell-q)^2-m_{\tilde{d}_s}^2\right)\delta\left((\ell-p)^2\right)
\label{integral5}\\
&\left.\left.\ \ \ \ \ \ \ \ + \int\frac{{\rm d}^4\ell}{(2\pi)^4}(-2q\cdot \ell + 2q\cdot p) (-i\pi)^2\delta\left(\ell^2 - |\mu|^2\right)P\frac{1}{(\ell-q)^2-m_{\tilde{d}_s}^2}\delta\left((\ell-p)^2\right)
\right\}\right]
\label{integral6}
\end{align}
 where $P$ indicates the principal value, $m_{\tilde{Q}_r},m_{\tilde{d}_s}$ respectively denote the soft SUSY breaking mass of isospin-doublet squark $\tilde{Q}_r$
 and isospin-singlet down-type squark $\tilde{d}_s$ ($r,s$ are flavor indices), $m_H$ denotes 
 the mass of a Higgs boson, and $p,\,q,\,p-q$ are respectively the four-momentum of the external squark, quark, and Higgsino.
Here the mass of the internal quark is neglected.
We choose the unitary gauge so that we do not need to consider Goldstone bosons propagating in the loop.
Still, the multiple physical Higgs bosons of MSSM, which have different masses and couplings, propagate in the loop, and we sum over them.
This summation is understood implicitly in the above expression.
Each integral is calculated as 
\begin{align}
(\ref{integral1})&=-\frac{m_H^2}{32\pi(m_{\tilde{u}_i}^2-|\mu|^2)}\log\left|\frac{ \left\{(m_{\tilde{Q}_r}^2-|\mu|^2)(m_{\tilde{u}_i}^2-|\mu|^2) + |\mu|^2m_H^2\right\} M_{u_l}^2   }{ m_H^2 (m_{\tilde{u}_i}^2-|\mu|^2)^2 }\right|,
\\
(\ref{integral2})&=-\frac{1}{32\pi(m_{\tilde{u}_i}^2-|\mu|^2)|\mu|^2}
\left\{ \left|m_{\tilde{Q}_r}^2-|\mu|^2\right|(m_{\tilde{u}_i}^2-|\mu|^2) \right.
\nn\\
&\left. \ \ \ \ \ \ \ \ \ \ \ \ \ \ \ \ - {\rm Sgn}\left(m_{\tilde{Q}_r}^2-|\mu|^2\right) \ |\mu|^2m_H^2
\log\left|
1+\frac{(m_{\tilde{Q}_r}^2-|\mu|^2)(m_{\tilde{u}_i}^2-|\mu|^2) }{ |\mu|^2m_H^2 }
\right| \right\},
\\
(\ref{integral3})&=\frac{1}{32\pi (m_{\tilde{u}_i}^2-|\mu|^2) m_{\tilde{u}_i}^2}
\left\{ F(m_H,m_{\tilde{Q}_r},m_{\tilde{u}_i})(m_{\tilde{u}_i}^2-|\mu|^2)  \right.
\nn\\
&\left. \ \ \ \ \ \ \ \ \ \ \ \ \ \ \ \ - m_{\tilde{u}_i}^2m_H^2
\log\left|\frac{(m_{\tilde{u}_i}^2 - m_{\tilde{Q}_r}^2 + m_H^2) - F(m_H,m_{\tilde{Q}_r},m_{\tilde{u}_i}) - \frac{2m_{\tilde{u}_i}^2m_H^2}{m_{\tilde{u}_i}^2-|\mu|^2}
}{(m_{\tilde{u}_i}^2 - m_{\tilde{Q}_r}^2 + m_H^2) + F(m_H,m_{\tilde{Q}_r},m_{\tilde{u}_i}) - \frac{2m_{\tilde{u}_i}^2m_H^2}{m_{\tilde{u}_i}^2-|\mu|^2}}
\right| \right\},
\\
(\ref{integral4})&=\frac{m_{\tilde{u}_i}^2+m_{\tilde{d}_s}^2-2|\mu|^2}{32\pi(m_{\tilde{u}_i}^2-|\mu|^2)}{\rm Sgn}\left(m_{\tilde{d}_s}^2-|\mu|^2\right)
\log\left|\frac{ (m_{\tilde{u}_i}^2+m_{\tilde{d}_s}^2-2|\mu|^2) (m_{\tilde{u}_i}^2m_{\tilde{d}_s}^2-|\mu|^4) M_{u_l}^2 }{ (m_{\tilde{u}_i}^2-|\mu|^2)^2 (m_{\tilde{d}_s}^2-|\mu|^2)^2 }\right|,
\\
(\ref{integral5})&=\frac{1}{32\pi(m_{\tilde{u}_i}^2-|\mu|^2)|\mu|^2}
\left\{ \left|m_{\tilde{d}_s}^2-|\mu|^2\right|(m_{\tilde{u}_i}^2-|\mu|^2) \right.
\nn\\
&\left. \ \ \ \ \ \ \ \ \ \ -{\rm Sgn}\left(m_{\tilde{d}_s}^2-|\mu|^2\right) \ |\mu|^2(m_{\tilde{u}_i}^2+m_{\tilde{d}_s}^2-2|\mu|^2)
\log\left|
1+\frac{(m_{\tilde{d}_s}^2-|\mu|^2)(m_{\tilde{u}_i}^2-|\mu|^2) }{ |\mu|^2(m_{\tilde{u}_i}^2+m_{\tilde{d}_s}^2-2|\mu|^2) }
\right| \right\},
\\
(\ref{integral6})&=-\frac{1}{32\pi(m_{\tilde{u}_i}^2-|\mu|^2)m_{\tilde{u}_i}^2}
\left\{ (m_{\tilde{u}_i}^2-|\mu|^2)^2 +m_{\tilde{u}_i}^2(2|\mu|^2 - m_{\tilde{u}_i}^2 - m_{\tilde{d}_s}^2)
\log\left|
1-\frac{ (m_{\tilde{u}_i}^2-|\mu|^2)^2 }{ |\mu|^4-m_{\tilde{u}_i}^2m_{\tilde{d}_s}^2 }
\right| \right\},
\end{align}
 where $F(a,b,c)=\sqrt{|a^4+b^4+c^4-2a^2b^2-2b^2c^2-2c^2a^2|}$, Sgn stands for the sign function, and $M_{u_l}$ denotes the mass of the final-state up-type quark.
Here $M_{u_l}$ is neglected unless it appears in logarithm and gives a large contribution.

We define the CP-violation parameter for the $\phi$ decay, $\epsilon_\phi$, as 
\begin{align}
\epsilon_\phi &= \frac{1}{\Gamma_\phi}\sum_{l=1}^3\frac{1}{6}\left\{
\Gamma(\tilde{u}^c_i \to Q_l^{u\dagger} \tilde{H}^{0\dagger})
+\Gamma(\tilde{d}^c_j \to Q_l^{d\dagger} \tilde{H}^{0\dagger})
+\Gamma(\tilde{d}^c_k \to Q_l^{d\dagger} \tilde{H}^{0\dagger}) \right.
\nn\\
&\ \ \ \ \ \ \ \ \ \ \ + \Gamma(\tilde{u}^c_i \to Q_l^{d\dagger} \tilde{H}^-)
+\Gamma(\tilde{d}^c_j \to Q_l^{u\dagger} \tilde{H}^+)
+\Gamma(\tilde{d}^c_k \to Q_l^{u\dagger} \tilde{H}^+)
\nn\\
&\ \ \ \ \ \ \ \ \ \ \ - \Gamma(\tilde{u}^{c\dagger}_i \to Q_l^u \tilde{H}^0)
-\Gamma(\tilde{d}^{c\dagger}_j \to Q_l^d \tilde{H}^0)
-\Gamma(\tilde{d}^{c\dagger}_k \to Q_l^d \tilde{H}^0)
\nn\\
&\left. \ \ \ \ \ \ \ \ \ \ \ - \Gamma(\tilde{u}^{c\dagger}_i \to Q_l^d \tilde{H}^+)
-\Gamma(\tilde{d}^{c\dagger}_j \to Q_l^u \tilde{H}^-)
-\Gamma(\tilde{d}^{c\dagger}_k \to Q_l^u \tilde{H}^-) \right\},
 \label{epsilondef}
\end{align}
 where $\Gamma_\phi$ is the total width given in Eq.~(\ref{totalwidth}).

The decay of $\phi$ non-thermally generates baryon number of the Universe.
The baryon number yield from $\phi$ decay, $n_B/s|_{{\rm from}\,\phi\, {\rm decay}}$, is given by
\begin{align}
\left.\left. \frac{n_B}{s}\right|_{{\rm from}\,\phi\, {\rm decay}} = -\frac{1}{3}\epsilon_\phi \frac{n_\Phi}{s}\right|_{{\rm at} \, \phi \, {\rm decay}},
\label{baryonyield}
\end{align}
 where $n_B$ denotes baryon number density, $s$ entropy density, and $n_\Phi$ the number density of the particle corresponding to the flat direction $\Phi$.
Here $n_\Phi/s$ at the time of $\phi$ decay is expressed with the reheating temperature $T_R$ and the mass term of the flat direction $m_\Phi$ as
\begin{align}
\left.\frac{n_\Phi}{s}\right|_{{\rm at} \, \phi \, {\rm decay}} = \frac{3}{4}\frac{g_{\rm eff}}{g_{\rm eff,S}}\frac{T_R}{m_\Phi},
\end{align}
 where $g_{\rm eff,S}$ is the effective entropy degree of freedom at temperature $T=T_R$.
The sphaleron process alters the baryon number yield and the value at present is given by~\cite{Khlebnikov:1988sr,Harvey:1990qw}
\begin{align}
\left. \frac{n_B}{s}\right|_{{\rm present}} = \frac{8}{23} \left. \frac{n_B}{s}\right|_{{\rm from}\,\phi\, {\rm decay}}.
\end{align}
\\

\section{Numerical analysis}

We present a benchmark parameter set that successfully explains the observed power spectrum amplitude $P_\zeta(k_*)$
 and the scalar spectral index $n_s$, satisfies the bound on tensor-to-scalar ratio $r$,
 and further explains the observed baryon number density through non-thermal baryogenesis discussed in Section~\ref{nonthermal}.

The benchmark is given as follows:
The flat direction is assumed to consist of isospin-singlet up, down and strange squarks, i.e., $i=1$, $j=1$ and $k=2$.
Also, the relevant parameters take the following values:
\begin{align}
&\lambda = 3.87,
\nn\\
&m_{\tilde{u}_1} = m_{\tilde{u}_2} = m_{\tilde{u}_3} = m_{\tilde{d}_1} = m_{\tilde{d}_2} = m_{\tilde{d}_3} = 3000~{\rm GeV},
\nn\\
&m_{\tilde{Q}_1} = m_{\tilde{Q}_2} = m_{\tilde{Q}_3} = 4000~{\rm GeV},
\nn\\
&M_{\tilde{g}}=2000~{\rm GeV},
\nn\\
&\mu = e^{-i\,0.0000850}\cdot 2000~{\rm GeV},
\nn\\
&A^u = A^d = 0,
\nn\\
&m_{H^\pm} = m_{H^0} = m_A = 4000~{\rm GeV},
\nn\\
&\tan\beta=3,
\nn\\
&\frac{\phi_*-\phi_0}{\phi_0} = -1.07\times 10^{-11},
\end{align}
 where $m_{H^\pm},m_{H^0},m_A$ respectively denote the masses of the charged, heavy CP-even, and CP-odd Higgs bosons.
The Yukawa coupling constants $y^u,y^d$ and the QCD gauge coupling $g_s$ are computed from experimental data as in Ref.~\cite{Haba:2022myj} at the renormalization scale of 2000~GeV.
It is easy to confirm that Eq.~(\ref{condition2}) holds with the above parameter values.
Therefore, the number of e-folds since the pivot scale exited the horizon is calculated with Eq.~(\ref{reheating}) along with Eq.~(\ref{hubble}) and Eq.~(\ref{nefoldspivot}) in Appendix~A.

For the above benchmark, the cosmological observables are predicted as
\begin{align}
P_\zeta(k_*)&=2.1\times 10^{-9},
\nn\\
n_s &= 0.9649,
\nn\\
r &= 9.3\times 10^{-31},
\nn\\
\left. \frac{n_B}{s}\right|_{{\rm present}}  &= 8.7\times 10^{-11}.
\end{align}
The above predictions are all consistent with the current data~\cite{Planck:2018jri,Tristram:2021tvh}.
This proves that our scenario of inflection point inflation and non-thermal baryogenesis that utilizes the MSSM $udd$ flat direction is viable.

We discuss the reheating temperature and its implications.
For the above benchmark, the reheating temperature $T_R$ is found to be
\begin{align}
T_R = 3.4\times 10^8~{\rm GeV}.
\end{align}
With such a high reheating temperature, the gravitino problem~\cite{Khlopov:1984pf,Ellis:1984eq} can occur.
However, severity of the gravitino problem depends on the soft SUSY breaking mechanism that we do not specify,
 and so this issue is beyond the scope of the present paper.

Since the reheating temperature is much higher than the soft SUSY breaking masses, 
 the MSSM particles, including the one-particle states of squark fields that have constituted the inflaton $\phi$, reach thermal equilibrium after the reheating.
Because the interactions of MSSM particles conserve baryon and lepton numbers, the baryon number yield created from the $\phi$ decay
 is not washed out by these interactions.
The difference between the interactions of MSSM particles that conserve baryon and lepton numbers, and the decay of the inflaton $\phi$ that creates baryon number,
 comes from the fact that $\phi$ is a superposition of a squark field and an antisquark field.
As a result, $\phi$ decays into both a quark$+$gaugino/Higgsino and an antiquark$+$gaugino/Higgsino,
 and asymmetry in the partial widths of the quark decay mode and the antiquark decay mode induces non-zero baryon number.
Conversely, such a superposition state cannot be created from the interactions of MSSM particles, and it should be prepared as an initial condition.
\\

\section{Conclusion}

We have proposed a scenario where a $udd$ flat direction of MSSM serves as the inflaton of inflection point inflation,
 and its decay at the reheating directly generates baryon asymmetry of the Universe.
We have derived the expressions for the cosmological parameters, and further calculated the CP asymmetry parameter and the baryon number yield for the decay of the inflaton.
We have confirmed that this scenario is compatible with the Planck/BICEP data on $P_\zeta (k_*)$, $n_s$ and the constraint on $r$, and successfully explains the observed baryon number density of the Universe.

\section{Acknowledgment}
This work is partially supported by Scientific Grants by the Ministry of Education, Culture,
Sports, Science and Technology of Japan, No. 23H03392 (NH) and No. 19K147101 (TY).

\section*{Appendix A}

In order to solve the horizon problem, the total number of e-folds $N_{\rm total}$ should satisfy the following bound:
 \begin{align}
 \frac{1}{H_{\rm inf}} e^{N_{\rm total}} \frac{a(t_{\rm rh})}{a(t_{\rm end})}\frac{a_0}{a(t_{\rm rh})} \ > \ \frac{1}{H_0},
 \end{align}
 where $H_{\rm inf}$ is the Hubble rate during inflation, $t_{\rm end},t_{\rm rh}$ respectively denote
 the time at the end of inflation and at the reheating, $a_0$ is the scale factor at present, and $H_0=67$~km/s/Mpc is the Hubble rate at present~\cite{PDG}.
If the entropy is conserved from $t=t_{\rm rh}$ to the present, we have
 $a_0/a(t_{\rm rh})=(g_{\rm eff}T_R^3/g_{S,{\rm eff},0}T_0^3)^{1/3}$, where $T_R$ is the reheating temperature,
 $g_{\rm eff}$ is the effective relativistic degree of freedom at the reheating,
 $T_0=2.73$~K is the CMB temperature at present, and $g_{S,{\rm eff},0}=43/11$ is the effective entropy degree of freedom at present.
By inserting the above values, the condition for solving the horizon problem is recast into
\begin{align}
N_{\rm total} > 68 - \log\frac{a(t_{\rm rh})}{a(t_{\rm end})} + \log\frac{H_{\rm inf}}{1~{\rm GeV}} - \log\frac{T_R}{1~{\rm GeV}} - \frac{1}{3}\log g_{\rm eff}.
\label{nefoldstotal}
\end{align}
Here $a(t_{\rm rh})/a(t_{\rm end})$ depends on details of the reheating process.


The number of e-folds since the comoving scale $k_*$ exited the horizon until the end of inflation $N(\phi_*)$ satisfies
\begin{align}
H_{\rm inf} \simeq H(\phi_*)=\frac{k_*}{a(t_*)} = \frac{k_*}{a_0}\frac{a_0}{a(t_{\rm rh})}\frac{a(t_{\rm rh})}{a(t_{\rm end})}e^{N(\phi_*)}.
\end{align}
For $k_*/a_0 = 0.05$~Mpc$^{-1}$, we get
\begin{align}
N(\phi_*) = 62 - \log\frac{a(t_{\rm rh})}{a(t_{\rm end})} + \log\frac{H_{\rm inf}}{1~{\rm GeV}} - \log\frac{T_R}{1~{\rm GeV}} - \frac{1}{3}\log g_{\rm eff}.
\label{nefoldspivot}
\end{align}
For this $k_*$, the condition for solving the horizon problem is re-expressed as
\begin{align}
N_{\rm total} > 6+N(\phi_*).
\end{align}


\end{document}